\documentstyle[12pt]{article}
\begin{document}
\thispagestyle{empty}
\vspace*{-2cm}
\begin{flushright}
{Alberta Thy-34-92}    \\
{CfPA-92-26}         \\
{December 1992}           \\
\end{flushright}
\vspace{1cm}
\begin{center}
{\bf \large
 QCD corrections to the charged Higgs decay\\ of  a  heavy  quark}
\end{center}
\vspace{1cm}
\begin{center}
{Andrzej Czarnecki}\\
\vspace{.3cm}
{\em Department of Physics, University of Alberta, Edmonton, Canada
T6G 2J1}
\end{center}
\begin{center}and\end{center}
\begin{center}
{Sacha Davidson}\\
\vspace{.3cm}
{\em Center for Particle Astrophysics, University of California
at Berkeley,\\ Berkeley, CA 94720, U.S.A.}
\end{center}
\hspace{3in}
\begin{abstract}
   We present an analytic formula for the $O(\alpha_s)$ corrections
to the decay $t\rightarrow H^+b$ for nonzero $m_H$ and $m_b$.
The Equivalence  Theorem is used to relate these corrections to those
for
 the process $t\rightarrow W^+b$. Contrary  to
previously published results we find that the effect of the $b$ quark
mass is negligible for a wide range of $m_t$, $m_H$ and $\tan \beta$.
\end{abstract}

\begin{flushleft}
PACS numbers: 12.15.cc, 12.38.Bx, 14.80.Dq
\end{flushleft}

\newpage

\section{Introduction}

 Many extensions of the Standard Model contain more than one Higgs
doublet.
The electroweak gauge bosons only absorb one of the charged Higgs
fields,
leaving the others as physical charged scalars, so that in a
two-doublet model the top could  decay to $H^+ b$ (if $m_t > m_{H^+}
+ m_b$).
For certain choices of parameters, this process dominates over the
expected
$t \rightarrow W^+ b$ decay, so it is of interest to correctly
calculate its
QCD corrections.

It is
well
known that the QCD corrections to the decay rate of a heavy quark
into a $W$
 are
of
 order  $10\%$ in the Standard Model~\cite{jk1,jk2}. Several groups
have undertaken  to
calculate these corrections for the decay into a charged
Higgs.  The
effect
of the soft gluons has been calculated in ref.~\cite{liuyao90}, and
the
decay $t\rightarrow H^+bg$, with real gluons only, has been studied
in
ref.~\cite{reid91}. The full one-loop QCD corrections have been
computed, at first by neglecting the mass of the $b$-quark in
ref.~\cite{liyuan90}, and then with a nonzero $m_b$ in
ref.~\cite{liyuan92} (in the framework of what is called the Model I
- see
discussion below). The claim of the latter paper is that the QCD
correction
for $m_t=150$~GeV, $m_b=4.5$~GeV and $\alpha_s=0.1$ is as large as
$-15\%$
in the (unphysical) limit of  the  massless Higgs. However, as will
be argued
in the following section,  according to the Equivalence Theorem (see
\cite{et1,et2,et3} as well as \cite{liuyao92} and references therein)
the
correction in this limit should be the same as the
correction to the decay $t \rightarrow W^+ b$, which for the above
values
of parameters is $-8.6\%$ \cite{jk1,cza90}. The purpose of the
present paper is
to reevaluate the
first order QCD corrections with non-zero  $b$-quark mass.  We use
dimensional regularization for both the
ultraviolet and the infrared divergences, which leads to simpler
algebra than if one assigns a finite mass to the gluon, as was done
in~\cite{liyuan90,liyuan92,liuyao92}. Our result is consistent with
the
Equivalence Theorem.

In a model with two Higgs doublets and generic couplings to all the
quarks, it is difficult to avoid flavour-changing neutral currents.
We
therefore
limit ourselves to models that naturally side-step these problems by
restricting the Higgs couplings \cite{hunter}.
 The first possibility is to have the
doublet $H_2$
coupling  to all the quarks, and the  $H_1$ doublet interacting with
none of
them. The
vacuum expectation value of  $H_1 = v_1$
 will nonetheless contribute to the $W$ mass, leading to an
$H^- t \bar{b}$ vertex of the form
\begin{equation}
\frac{g V_{tb} }{\sqrt{2} m_W} H^- \bar{b} \left\{ m_t \cot \beta \,
{\rm R} -
m_b  \cot \beta \, {\rm L} \right\} t
{}~~~({\rm model~I})     \label{s3}
\end{equation}
where $H^-$ is the physical charged Higgs, $V_{tb}$ is the `33'
element of
the CKM matrix,  L and R are the chiral projection
operators, and $\tan \beta = v_2/v_1$ is the ratio of the vacuum
expectation values of the two Higgs. The second possibility is to
have
$H_2$ couple to the right-handed up-type quarks ($u_R, c_R, t_R$),
and the  $H_1$ couple to the right-handed down-type quarks. This is
what happens in the Minimal Supersymmetric Standard Model. It is easy
to show that the interaction Lagrangian
\begin{equation}
H_2 \bar{u}_R^i h^{(u)}_{ij} q^j_L + H_1 \bar{d}_R^i h^{(d)}_{ij}
q_L^j + h.c.
\label{s1}
\end{equation}
leads to the vertex
\begin{equation}
\frac{g V_{tb} }{\sqrt{2} m_W} H^- \bar{b} \left\{ m_t \cot \beta \,
{\rm R} -
m_b  \tan \beta \, {\rm L} \right\} t
{}~~~({\rm model~II})     \label{s2}
\end{equation}
where we have numbered the models in accordance with \cite{hunter}.

In order to simplify the following formulas  we introduce
dimensionless
parameters for the scaled masses:
\begin{eqnarray}
\epsilon={m_b\over m_t}, \rule{.3in}{0in}
\chi={m_H\over m_t},  \rule{.3in}{0in}
w = {m_W\over m_t},
\label{eq:def}
\end{eqnarray}
and write the vertex $t\rightarrow H^+b$
as
\begin{eqnarray}
i{g\over 2\sqrt{2}w}V_{tb} \bar{b}
\left({\rm a}+{\rm b}\gamma_5\right)t.
\label{eqab}
\end{eqnarray}
where from (\ref{s3}) and (\ref{s2})
\begin{eqnarray}
{\rm Model\rule{0.1in}{0in}  I}:\rule{0.2in}{0in}
          \left\{ \begin{array}{c}
{\rm a}=\cot\beta(1-\epsilon)\\
{\rm b}=\cot\beta(1+\epsilon)
\end{array}             \right.  \rule{0.3in}{0in}
{\rm Model\rule{0.1in}{0in}  II}:\rule{0.2in}{0in}
          \left\{ \begin{array}{c}
{\rm a}=\cot\beta+\epsilon\tan\beta\\
{\rm b}=\cot\beta-\epsilon\tan\beta
\end{array}                   \right.\nonumber
\end{eqnarray}

The next section contains our result, which is examined in the
Discussion.
The Appendix contains some details of the calculation.

\section{QCD Corrections}
The notation we use is similar to that used in the analysis of
semileptonic
decays \cite{jk1,jk2}.
In terms of the dimensionless parameters (\ref{eq:def}),
we define the following kinematic variables:
\begin{eqnarray}
\bar P_0 &\equiv &{1\over 2}(1-\chi^2+\epsilon^2)\nonumber\\
\bar P_3 &\equiv &{1\over
2}\sqrt{1+\chi^4+\epsilon^4-2(\chi^2+\epsilon^2+\chi^2 \epsilon^2)}
\nonumber\\
\bar P_{\pm}&\equiv &\bar P_0\pm\bar P_3\nonumber\\
\bar Y_p &\equiv &{1\over 2}\ln{\bar P_+\over\bar P_-}\nonumber\\ &&
\nonumber\\
\bar W_0 &\equiv &{1\over 2}(1+\chi^2-\epsilon^2)\nonumber\\
\bar W_{\pm}&\equiv &\bar W_0\pm\bar P_3\nonumber\\
\bar Y_w &\equiv &{1\over 2}\ln{\bar W_+\over\bar W_-}
\end{eqnarray}

 The tree level decay rate is
\begin{equation}
\Gamma^0 (t \rightarrow H^+ b) = \frac{G_Fm_t^3}{4 \sqrt2 \pi} |
V_{tb}|^2
[\bar{P}_0 ({\rm a}^2 + {\rm b}^2) + \epsilon ({\rm a}^2 - {\rm
b}^2)]\bar{P}_3
\end{equation}
and
the $O(\alpha_s)$ correction is
\begin{eqnarray}
\Gamma^{(1)}={\alpha_s\over 6\pi^2}{G_F m^3_t \left|V_{tb}\right|^2
\over \sqrt{2}}\left[\left({\rm a}^2+{\rm b}^2\right)G_+
+\left({\rm a}^2-{\rm b}^2\right)\epsilon G_- +{\rm ab}G_0\right]
\label{eq:main}
\end{eqnarray}
with
\begin{eqnarray}
 G_+&=&\bar P_0
 {\cal H}+{\bar P_0}\bar P_3 \left[{9\over 2}-4\ln\left({4 {\bar
P_3}^2\over\epsilon\chi} \right)\right]
\nonumber\\ &&
  +{1\over 4\chi^2} {\bar Y_p} \left(2-\chi^2-4\chi^4
     +3\chi^6-2\epsilon^2-2\epsilon^4+2\epsilon^6-4\chi^2\epsilon^2-5
\chi^2\epsilon^4\right),
     \nonumber\\
      G_-&=&{\cal H}+{\bar P_3} \left[6-4\ln\left({4 {\bar
P_3}^2\over\epsilon\chi} \right)\right]
  +{1\over \chi^2} {\bar Y_p} \left(1-\chi^2-2\epsilon^2+\epsilon^4
      -3\chi^2\epsilon^2\right),
\nonumber\\
G_0&=&     -6{\bar P_0}{\bar P_3}\ln\epsilon,
\end{eqnarray}
and
\begin{eqnarray}
\lefteqn{ {\cal H}=4\bar P_0\left[{\rm Li}_2\left({{\bar
P_+}}\right)-{\rm
Li}_2\left({P_-}\right)-2{\rm Li}_2\left({1-{{\bar P_-}\over{\bar
P_+}}}\right)
\right.}
\nonumber\\ && \left.
+{\bar Y_p}\ln\left({{4{\bar P_3}^2\chi\over{\bar
P_+}^2}}\right)-{\bar
Y_w}\ln\epsilon\right]+
2\bar Y_w\left(1-\epsilon^2\right)+{2\over \chi^2}\bar
P_3\ln{\epsilon}
\left(1+\chi^2-\epsilon^2\right). \nonumber
\end{eqnarray}
In the limit of the zero mass of the $b$ quark the QCD correction
becomes
\begin{eqnarray}
\lim_{\epsilon\rightarrow 0}\Gamma^{(1)}={\alpha_s\over 6\pi^2}
{G_F m_t^3 \left| V_{tb}\right|^2\over\sqrt{2}}\cot^2\beta
     \left(2\tilde{G}_++\tilde{G}_0\right),
\label{eq:beta0}
\end{eqnarray}
where
\begin{eqnarray}
\tilde{G}_+&=&\left( 1-\chi^2\right)^2\left[{\rm
Li}_2\left({1-\chi^2}\right)
-{\chi^2\over 1-\chi^2}\ln\chi \right.\nonumber\\ && \left.
+\ln\chi\ln\left(1-\chi^2\right)+{1\over
2\chi^2}\left(1-{5\over 2}\chi^2\right)\ln\left(1-\chi^2\right)
-{\pi^2\over 3}+{9\over 8}+{3\over 4}\ln\epsilon
\right],       \nonumber\\
\tilde{G}_0&=&-{3\over 2}\left( 1-\chi^2\right)^2\ln\epsilon,
\end{eqnarray}
and we see that the mass singularities $\sim\ln\epsilon$ cancel in
the
expression
for the total rate (\ref{eq:beta0}). Our result in this limit is
identical
to the one obtained by Liu and Yao \cite{liuyao92} and is in
agreement with the
corrected version of ref.~\cite{liyuan90}.

If we further take the limit $m_H\rightarrow 0$ the rate becomes:
\begin{eqnarray}
\lim_{\epsilon,\chi\rightarrow 0}\Gamma^{(1)}=
{\alpha_s\over 6\pi^2}
{G_F m_t^3 \left| V_{tb}\right|^2\over\sqrt{2}}\cot^2\beta
     \left( {5\over 4}-{\pi^2\over 3}\right),
\end{eqnarray}
which is in agreement with the conclusion of the ref.~\cite{liuyao92}
as well as with our previous result \cite{sa92a}.

Now we would like to compare the corrections to the decay width
\linebreak
$\Gamma\left(t\rightarrow H^+b\right)$ with those to
$\Gamma\left(t\rightarrow W^+b\right)$. For simplicity we now take
$m_b=0$ and
$\cot\beta=1$, and examine the ratio of the first order correction to
the Born
rate:
\begin{eqnarray}
f_H(\chi)&=&{\Gamma^{(1)}\left(t\rightarrow H^+b\right)\over
\Gamma^{(0)}\left(t\rightarrow H^+b\right)}, \nonumber\\
f_W(w)&=&{ \Gamma^{(1)}\left(t\rightarrow W^+b\right)\over
\Gamma^{(0)}\left(t\rightarrow W^+b\right)}.
\end{eqnarray}
It has been noted in \cite{liuyao92} that in the limit of the
infinite top mass
these ratios are equal: $f_H(0)=f_W(0)$. On the other hand, when
$m_{H}$
approaches  $m_{t}$, we have:
\begin{eqnarray}
f_H(\chi) \stackrel{\chi\rightarrow 1}{\longrightarrow}
{\alpha_s\over 3\pi}
\left[-6\ln(1-\chi^2)-{8\over 3}\pi^2+13\right].
\end{eqnarray}
By comparison with \cite{jk1} we see that:
\begin{eqnarray}
\lim_{x \rightarrow 1}{f_H(x)\over f_W(x)}=1.
\end{eqnarray}

Finally, we examine the corrections in the limiting case  where the
mass of the
charged Higgs is zero but the $b$ quark mass  is finite. This is of
course
unphysical, but serves as a useful check on our equations. If we
choose the
parameters a and b from (\ref{eqab}) to correspond to the couplings
of the
single Standard Model Higgs, then the Equivalence Theorem implies
that
the  corrections are the same
as in the process $t\rightarrow bW$ in the limit of  massless
$W$ boson and nonzero $m_b$. The latter can be obtained by taking the
limit of the relevant formula \cite{jk1,cza90}:
\begin{eqnarray}
\lim_{m_W\rightarrow 0}\Gamma^{(1)}\left(t\rightarrow bW\right)&
=&{\alpha_s\over24\pi^2}{G_Fm_t^3\left|V_{tb}\right|^2\over
\sqrt{2}}\nonumber\\
&&\rule{-25mm}{0mm}
\left\{8(1-\epsilon^2)^2(1+\epsilon^2)
\left[{\rm Li_2}\left(\epsilon^2\right)-{\pi^2\over
6}+\ln\left(\epsilon\right)\ln\left(1-\epsilon^2\right)\right]
\nonumber
\right.\\&&\left.\rule{-25mm}{0mm}
-4\epsilon^2\left(7-5\epsilon^2+4\epsilon^4\right)\ln\left(\epsilon
\right)
-8\left(1-\epsilon^2\right)^3\ln\left(1-\epsilon^2\right)
\nonumber\right.\\&& \left.
-(1-\epsilon^2)(-5+22\epsilon^2-5\epsilon^4)\rule{0mm}{6mm}\right\}
\label{eq:twb}
\end{eqnarray}
The couplings of the Goldstone boson charged Higgs of the Standard
Model
(longitudinal $W$) to $t$ and $b$ can easily be calculated to be
those of
Model I, with $\cot \beta = 1$.
In  this case, the  corrections to the decay $t\rightarrow bH^+$ are,
in
the  limit $m_H \rightarrow 0$:

\begin{eqnarray}
\lim_{m_H\rightarrow 0}\Gamma^{(1)}\left(t\rightarrow bH^+\right)&
=&{\alpha_s\over 6\pi^2}{G_F m^3_t \left|V_{tb}\right|^2
\over \sqrt{2}}\nonumber\\ &&
\left[2(1+\epsilon^2)G_+^0-4\epsilon^2G_-^0+(1-\epsilon^2)G_0^0\right
]
\label{eq:mai}
\end{eqnarray}
where $G_i^0$ are limits of corresponding functions $G_i$ for
$m_H= 0$: \newpage
\begin{eqnarray}
G_+^0&=&(1+\epsilon^2)^2
\left[{\rm Li_2}\left(\epsilon^2\right)-{\pi^2\over
6}+\ln\left(\epsilon\right)\ln\left(1-\epsilon^2\right)\right]
\nonumber\\
&& +\left(
{3\over4}+\epsilon^2-{5\over4}\epsilon^4\right)\ln\left(\epsilon
\right)     -\left(1-\epsilon^4\right)\ln\left(1-\epsilon^2\right)
+{5\over 8}(1-\epsilon^4) \nonumber\\
G_-^0&=&2(1+\epsilon^2)
\left[{\rm Li_2}\left(\epsilon^2\right)-{\pi^2\over
6}+\ln\left(\epsilon\right)\ln\left(1-\epsilon^2\right)\right]
\nonumber\\
&& +\left(3-\epsilon^2\right)\ln\left(\epsilon\right)
  -2\left(1-\epsilon^2\right)\ln\left(1-\epsilon^2\right)
+2(1-\epsilon^2) \nonumber\\
G_0^0&=&-{3\over 2}(1-\epsilon^4)\ln(\epsilon)
\end{eqnarray}
Inserting these expressions into equation (\ref{eq:mai}) we obtain
 the same formula as
(\ref{eq:twb}).

\section{Discussion}

In  Figure 1 the ratio of the first order QCD correction to the Born
rate for  the decay $t\rightarrow H^+b$ is plotted as the function of
the
ratio of masses $\chi=m_H/m_t$. We have chosen the set of
 parameters $\cot\beta=1$, {$m_t=150$~GeV} for an easy comparison
with the
analogous diagram in ref.~\cite{liyuan92}. It can be seen that the
effect of the mass of the $b$-quark is negligible, except in the case
of
$m_t-m_H\sim m_b$.

There are also logarithmic corrections  ($\sim \epsilon^2  \ln
\epsilon$) to
the
decay rate in model II, as can be seen from Figure 2.
Here we compare the branching ratios of the
decays $t\rightarrow H^+b$ and $t\rightarrow W^+b$, taking
{$m_t=100~$GeV}
so that this plot can be easily compared with a similar one
 published in ref.~\cite{hunter}. Our graph is different from theirs
 in that we now include  QCD corrections  to both decay rates. These
corrections
modify the diagram significantly only in the case of large values of
$\tan\beta$ in  Model II.
It must be noted however, that although the corrections are
relatively
large, the top decays principally to $W^+ b$ in this region of $\tan
\beta$.
In model I, both a and b are proportional to $\cot \beta$, so the log
of the branching ratio as a function of $\tan \beta$ decreases with a
slope of
-2. As can be seen from figure 2, the $\ln \epsilon$ corrections
cancel to
order $\epsilon^2$ among $G_+, G_-$ and $G_0$. However in model II,
the
decay rate is a polynomial in $\tan \beta$ with exponents -2, 0 and
2,
and the $\ln \epsilon$ does not cancel in the ``0'' and ``2'' terms.

\section*{Summary}

We have calculated an analytic expression for the $O(\alpha_s)$
corrections to the decay $ t \rightarrow H^+ b$ for non-zero $m_H$
and
$m_b$. The $m_b = 0$ limit of our result is the same as in
\cite{liuyao92} and
the corrected version of \cite{liyuan90};
however, our full expression disagrees with the $m_b \neq 0 $ result
presented in \cite{liyuan92}. To check the $m_b$ dependence of our
results we
have compared it with the corresponding formula for the decay
$t\rightarrow
W^+b$.

\section*{Acknowledgments}
We would like to thank Professor Bruce Campbell, Professor M. K.
Gaillard,
 and Professor
A.~N.~Kamal for interest in this work and  important discussions and
comments.
A.C. gratefully acknowledges support from the Killam Foundation;
S.D.'s
research is partially supported by NSERC. This
research was also partially supported by a grant to A.~N.~Kamal from
the
Natural Sciences and Engineering Council (NSERC) of Canada.
 \appendix
\section{Details of the calculation}
Throughout this calculation we have used dimensional regularization,
working in $d=4-2\omega$ dimensions. The counterterm for the vertex
function has been calculated
according to ref.~\cite{liyuan90,bl80}:
\begin{eqnarray}
\delta\Lambda &=& {\rm a}\left\{{1\over 2}(Z_t-1)+{1\over
2}(Z_b-1)\right\}
-{{\rm a}+{\rm b}\over 2}{\delta m_t\over m_t}
-{{\rm a}-{\rm b}\over 2}{\delta m_b\over m_b} \nonumber\\
&&+\left[{\rm b}\left\{{1\over 2}(Z_t-1)+{1\over 2}(Z_b-1)\right\}
-{{\rm a}+{\rm b}\over 2}{\delta m_t\over m_t}
+{{\rm a}-{\rm b}\over 2}{\delta m_b\over m_b} \right]\gamma_5
\nonumber
\end{eqnarray}
and the renormalization constants are (we take the renormalization
scale
equal to the mass of the decaying quark):
\begin{eqnarray}
&&Z_t-1=-{\delta m_t\over m_t}={\alpha_s\over
3\pi}\left(-{3\over\omega}
+3\gamma-3\ln{4\pi}-4\right),\nonumber\\
&&Z_b-1=-{\delta m_b\over m_b}={\alpha_s\over
3\pi}\left(-{3\over\omega}
+3\gamma-3\ln{4\pi\over\epsilon^2}-4\right).
\end{eqnarray}
The contribution of the virtual corrections to the total decay rate
is:
\begin{eqnarray}
\Gamma^{(1)}_{virt}={\alpha_s\over 6\pi^2}{G_F m^3_t
\left|V_{tb}\right|^2
\over \sqrt{2}} \left[\left({\rm a}^2+{\rm b}^2\right)V_+
+\left({\rm a}^2-{\rm b}^2\right)\epsilon V_- +{\rm ab}G_0\right],
\label{eq:virt}
\end{eqnarray}
where
\begin{eqnarray}
V_+&=&\bar P_0{\cal V}
+{\bar Y_p\over 2\chi^2}\left[(1-\epsilon^2)^2(1+\epsilon^2)
-\chi^4(3+3\epsilon^2-2\chi^2)\right],\nonumber\\
V_-&=&{\cal V}+\bar Y_p{(1-\epsilon^2)^2-\chi^4\over\chi^2},
\end{eqnarray}
with
\begin{eqnarray}
\cal V&=&
2\bar P_3
\left[-{1\over\omega}+2\gamma+2\ln{\bar P_3
\over 2\pi}-3+{1\over\chi^2}(1-\epsilon^2+\chi^2)\ln\epsilon\right]
\nonumber \\
&+&2\bar P_0
\left[{\rm Li_2}\left(\bar P_+\right)-{\rm Li_2}\left(\bar P_-\right)
-{\rm Li_2}\left(1-{\bar P_-\over\bar P_+}\right)-\bar
Y_p^2-2\ln\epsilon\bar
Y_w\right. \nonumber \\
&&\left. -2\bar Y_p \left(-{1\over 2\omega}+\gamma+\ln{\bar
P_3\epsilon\over 2\pi\chi}\right)\right].\nonumber
\end{eqnarray}

 The real gluon radiation  calculation requires an
integration over  three body phase space $\Phi(t;H,b,g)$. This leads
to divergences due to the emission of soft and collinear gluons, for
which we
use dimensional regularization (see ref.~\cite{xypham} and
references quoted therein).
 The phase space integration we are considering now
is analogous to the decay $t\rightarrow Wbg$, for which  the relevant
integrals have been listed in \cite{cza90}. Adding the result to
the virtual gluon contribution (\ref{eq:virt}) yields the final
formula
(\ref{eq:main}).

We would like to add that this calculation was greatly facilitated by
the use of algebraic manipulation programs FORM~\cite{form} and
Mathematica~\cite{wolf}.

 \newpage
\begin{figure}
\begin{picture}(0,100)(430,260)
\put(440,420){${\Gamma^{(1)}\over\Gamma^{(0)}}$}
\put(750,270){${m_H\over m_t}$}
\end{picture}
\caption{Ratio of the first order correction to the Born rate for
$\tan\beta=1$
and $m_t=150$~GeV: $m_b=4.5$ GeV
(Model I - solid, Model II - dotted), $m_b=0$ (dash-dotted line). }
\end{figure}

\begin{figure}
\begin{picture}(0,250)(430,260)
\put(405,470){${BR(t\rightarrow bH^+)\over BR(t\rightarrow bW^+)}$}
\put(750,270){${\tan\beta}$}
\end{picture}
\caption{Ratio of the branching ratios of $t\rightarrow bH^+$ and
$t\rightarrow bW^+$ for $m_b=4.5$ GeV: in the Model I with and
without QCD
corrections (solid  and dashed lines, undistinguishable) and in the
Model II
(dash-dotted  and dotted lines, respectively).}
\end{figure}

\end{document}